# Multi-TE Single-Quantum Sodium ($^{23}$Na) MRI: A Clinically Translatable Technique for Separation of Mono- and Bi-T$_2$ Sodium Signals

***Short title:*** Mono- and bi-T$_2$ sodium separation

Yongxian Qian,[1*] Ying-Chia Lin,[1] Xingye Chen,[1,2] Yulin Ge,[1] Yvonne W. Lui,[1,3] Fernando E. Boada[1†]

[1] Bernard and Irene Schwartz Center for Biomedical Imaging, Department of Radiology, New York University Grossman School of Medicine, New York, NY 10016.

[2] Vilcek Institute of Graduate Biomedical Sciences, NYU Grossman School of Medicine, New York, NY 10016.

[3] Department of Radiology, NYU Langone Health, New York, NY 10016.

**\* Corresponding author**. Email: Yongxian.Qian@nyulangone.org

† Present address: Department of Radiology, Stanford University, Stanford, CA 94305.

**Abstract** (194 words)

Sodium magnetic resonance imaging (MRI) is sensitive and specific to ionic balance of cells owing to 10-fold difference in sodium concentration across membrane actively maintained by sodium-potassium (Na$^+$-K$^+$) pump. Disruption of the pump and/or membrane integrity, as seen in neurological disorders such as epilepsy, multiple sclerosis, bipolar disease, and mild traumatic brain injury, leads to a large increase in intracellular sodium. Such a cellular-level alteration is however overshadowed by large signal from extracellular sodium, leaving behind a long-standing pursuit to separate signals from sodium exhibiting mono- vs. bi-exponential transverse (T$_2$) decay under the inherent constraint of low signal-to-noise ratio even at advanced clinical field of 3 Tesla. Here we propose a novel technique that exploits intrinsic difference in their T$_2$ decays by simply acquiring single-quantum images at multiple echo times (TEs) and performing accurate matrix inversion at voxel. This approach was then investigated using numerical models, agar phantoms and human subjects, showing high accuracy of the separation in phantoms (95.8% for mono-T$_2$ and 72.5–80.4% for bi-T$_2$) and clinical feasibility in humans. Thus, sodium MRI at 3T can now facilitate detection of neurological disorders early at cellular level and response to treatment as well.

## Acknowledgments

The authors would like to thank Dr. Ryan Brown and Mr. Karthik Lakshmanan for their support in the development of the 8-channel array sodium coil, Dr. Timothy Shepherd for his support in the recruitment of epilepsy patients, and Dr. Tiejun Zhao for his support and discussion in sodium MRI.

## Funding:

National Institutes of Health (NIH) grants RF1/R01 AG067502 (YQ, YL, XC, YG, YWL, FEB), R01 NS131458 (YQ, YWL, FEB), RF1 NS110041 (YG), R01 AG077422 (YG), R01 NS113517 (YQ, FEB), R01 NS108491 (YQ, FEB), and R01 CA111996 (YQ, FEB).

Department of Radiology of the New York University Grossman School of Medicine General Research Fund (YQ, YG, YWL, FEB).

## NIH-sponsor required statement:

Research reported in this publication was supported in part by the National Institute On Aging (NIA) of the National Institutes of Health (NIH) under Award Number RF1/R01AG067502. The content is solely the responsibility of the authors and does not necessarily represent the official views of the National Institutes of Health.

This work was performed under the rubric of the Center for Advanced Imaging Innovation and Research (CAI$^2$R), a National Institute of Biomedical Imaging and Bioengineering (NIBIB) Biomedical Technology Resource Center grant NIH P41 EB017183.

## Author contributions:
Conceptualization: YQ, FEB
Methodology: YQ, YWL, FEB
Investigation: YQ, YL, XC, YG, YWL, FEB
Visualization: YQ, YWL, FEB
Supervision: FEB
Writing – original draft: YQ
Writing – review & editing: YQ, YL, XC, YG, YWL, FEB

**Competing interests:**

YQ and FEB are inventor of the U.S. Patent Application, No.: 63/591,751. Title: *Process of multiple echo time acquisitions for quantitative bi-$T_2$ sodium magnetic resonance imaging*. Filed on October 19, 2023. IP Owner: New York University.

All other authors declare they have no competing interests.

This work was partially presented in the 25th Annual Meeting of ISMRM 2017, Honolulu, Hawaii, USA.

This work was presented in a different version on the preprint platform arXiv.com (https://arxiv.org/abs/2407.09868).

**Data and materials availability:**
Upon written request to the corresponding author, Yongxian Qian, PhD, all data (sodium images and FID signals) and codes used in this work (main text and supplementary materials) are available to any researcher solely for scientific research purposes.

**Introduction** (741 words)

In human brains, sodium ions (Na$^+$), when exposed to an electric field gradient of negatively charged macromolecules and proteins, experience nuclear quadrupolar interaction that results in bi-exponential decay in transverse (T$_2$) relaxation of nuclear spins when ions are not in fast motion – a situation in which correlation time between sodium ions and electric field gradient is much shorter than the inverse of Larmor frequency, $\tau_c \ll 1/\omega_0$ (*1, 2*). On the other hand, sodium ions in relatively fast motion cancel out the effect of quadrupolar interactions, resulting in mono-exponential T$_2$ decay (*1–4*). Historically, short-T$_2$ component of the bi-T$_2$ decay was mis-considered arising from "bound" sodium (*5, 6*) as it was not detectable by then-NMR (nuclear magnetic resonance) (*1–4*). The terminology of "bound sodium" however remains in use in today's literature of sodium MRI, but it comes into question more recently (*7*). For clarity, this article refers to "bi-T$_2$" sodium as those showing bi-exponential T$_2$ decay and "mono-T$_2$" sodium as those showing mono-exponential T$_2$ decay. Of note, mono- and bi-T$_2$ sodium ions can appear in both intra- and extracellular spaces (*8–10*), dependent on relative correlation time with electric field gradient (*1, 2*).

Sodium ($^{23}$Na) MRI (magnetic resonance imaging) currently acquires signals from both mono- and bi-T$_2$ sodium ions, and quantifies total sodium concentration (TSC) at voxels of an image. TSC is unique measure for non-invasive assessment of disruption in ionic homeostasis of cells in, or recovery from, pathological conditions including stroke, tumor, multiple scleroses, epilepsy, bipolar disorder, and mild traumatic brain injury (*8–12*). However, TSC is dominated by mono-T$_2$ sodium in cerebrospinal fluid (CSF) which has a high sodium concentration (~145 mM) and overshadows alteration in intracellular sodium which has a much lower concentration (~15 mM). Separation of mono- and bi-T$_2$ sodium signals can remove CSF impact and highlight intracellular alterations, especially at early stage of a disease happening at cellular level or in early (cellular) response to a treatment.

The difference in T$_2$ relaxation was extensively explored in sodium MRI as a means to separate the two populations of sodium ions in the brain. Triple quantum filtering (TQF) was considered a standard for human studies, in which magnetic resonance (MR) signals were generated solely from triple-quantum (TQ) transitions (*13*). TQF techniques however require multiple radiofrequency (RF) pulses for excitation and multi-step phase cycling to eliminate single-quantum (SQ) signals (*13–17*), leading to long scan time (20–40 min) and high specific absorption rate (SAR) causing a safety concern. More problematic is that TQF has much low signal-to-noise ratio (SNR) about 10 folds lower than SQ (*15–17*). These difficulties hamper TQF to be widely used on humans.

Alternative approaches were proposed. Inversion recovery (IR), adopted from proton ($^1$H) MRI, exploits a difference in longitudinal ($T_1$) relaxation between mono- and bi-$T_2$ sodium ions, and suppresses signals from mono-$T_2$ sodium of longer $T_1$ time (*18–20*). IR approach needs an extra RF pulse for the suppression and worsens SAR issue, unfavorable to human studies (*20, 21*). It also suffers from incomplete suppression of mono-$T_2$ sodium signals which are ~10 folds higher than bi-$T_2$ sodium signals, due to spatial inhomogeneity of $B_1^+$ field although adiabatic pulses are usually used, and complicates quantification of bi-$T_2$ sodium owing to unknown residual mono-$T_2$ sodium signals (*9,11,18*). To overcome these drawbacks, another alternative approach, called short-$T_2$ imaging, was proposed in which SQ images were acquired at multiple echo times (TEs) and then subtracted from each other to produce an image of the short-$T_2$ component of bi-$T_2$ sodium (*22–25*). In such a way, SAR was reduced to, and SNR was increased to, the level of SQ images, favorable to human studies in clinic. Unfortunately, the subtraction could not completely eliminate mono-$T_2$ sodium signal (~20% in residual), degrading accuracy of bi-$T_2$ sodium quantification (*25*).

In this study, the concept of short-$T_2$ imaging is generalized to multi-TE single-quantum (MSQ) imaging with a more accurate matrix inversion replacing the subtraction to substantially improve accuracy of the separation between mono- and bi-$T_2$ sodium signals. To develop MSQ technique, we optimized TEs for data acquisition, investigated impact of $T_2$ values on accuracy of the separation, and acquired free induction decay (FID) signals to generate $T_2^*$ spectrum for the matrix equation. To test MSQ technique, we implemented numerical simulations, phantom experiments, and human studies. The results are supportive of the proposed MSQ technique. We also itemized limitations of the MSQ technique and potential pitfalls in interpretation of the separated sodium signals.

**Results** (1,518 worlds)

*Model of sodium MRI signals*

A two-population model, Eq. (1), is used to describe single-quantum sodium image signal, *m(t)*, evolving with time *t* at an imaging voxel $\Delta V$. Time *t* = 0 is at the center of excitation RF pulse.

$$m(t) = m_{mo}\, Y_{mo}(t) + m_{bi}\, Y_{bi}(t), \ t \geq 0 \qquad (1)$$

With $m_{mo} \geq 0$, $m_{bi} \geq 0$, and $m_{mo} + m_{bi} = m(0)$

$$Y_{mo}(t) \equiv \exp(-t/T_{2,mo}) \qquad (1A)$$

$$Y_{bi}(t) \equiv a_{bs} \exp(-t/T_{2,bs}) + a_{bl} \exp(-t/T_{2,bl}) \qquad (1B)$$

The $m_{mo}$ and $m_{bi}$ are signal intensity proportional to volume fraction $v_q$ and sodium concentration $C_q$, i.e., $m_q \propto \Delta V v_q C_q$, $q$=*mo* and *bi*, for mono-T$_2$ and bi-T$_2$ sodium populations in a voxel $\Delta V$, respectively. $Y_{mo}(t)$ is relaxation decay of the mono-T$_2$ sodium of time constant $T_{2,mo}$ while $Y_{bi}(t)$ for the bi-T$_2$ sodium with $a_{bs} = 0.6$ for the short-T$_2$ component $T_{2,bs}$, and $a_{bl} = 0.4$ for the long-T$_2$ component $T_{2,bl}$. The split of 60:40 % in intensity of bi-T$_2$ sodium is from theoretical and experimental results for individual sodium nuclear spins (*1–4, 26*). These T$_2$ values are in an order of $T_{2,bs} \ll T_{2,bl} \leq T_{2,mo}$. However, Eq. (1) doesn't include mono-T$_2$ sodium of short T$_2$ value.

*Separation of mono- and bi-T$_2$ sodium signals*

Given SQ sodium images at multiple echo times, TEs = {$TE_1$, $TE_2$, …, $TE_N$}, in which TE is defined as time interval between the center of excitation RF pulse and center of the *k*-space, Eq. (1) becomes a matrix equation Eq. (2).

$$\mathbf{M} = \mathbf{Y}\mathbf{X} \qquad (2)$$

$$\mathbf{M} \equiv (m_1, m_2, \ldots, m_N)^T, \qquad m_i = m(TE_i), i = 1,2,\ldots,N \qquad (2A)$$

$$\mathbf{X} \equiv (m_{mo}, m_{bi})^T \qquad (2B)$$

$$\mathbf{Y} \equiv \begin{pmatrix} Y_{mo}(TE_1) & Y_{bi}(TE_1) \\ Y_{mo}(TE_2) & Y_{bi}(TE_2) \\ \vdots & \vdots \\ Y_{mo}(TE_N) & Y_{bi}(TE_N) \end{pmatrix} \qquad (2C)$$

Superscript *T* is operator for matrix transpose. A solution to Eq. (2) is given in Eq. (3) via an established algorithm called non-negative least-squares (NNLS) (*27*) in which the non-negative condition on **X** is incorporated into the solution.

$$\mathbf{X} = NNLS(\mathbf{Y}\mathbf{X} - \mathbf{M}) \qquad (3)$$

*Flowchart of the MSQ approach*

The proposed MSQ approach is illustrated in Fig. 1. The inputs are multi-TE sodium images and an FID signal. The outputs are mono-T$_2$, bi-T$_2$, and total sodium images, as well as maps of field inhomogeneity $\Delta B_0$ and single-term exponential fitted effective T$_2$ (called T$_2$*) – single-T$_2$* for short. Hereinafter, T$_2$* replaces T$_2$ as spin echo is not favorable to sodium MRI. Motion correction (MoCo) across multi-TE images is optional. Also optional is the low-pass (LP) filtering, which is a 3D averaging over a size of 3×3×3 voxels for instance, to reduce random noise on the bi-T$_2$ sodium images. The $\Delta B_0$ and single-T$_2$* maps present spatial distributions of the $B_0$ field inhomogeneity and the short- and long-T$_2$* components, providing indications for uncertain short-

$T_2^*$ decays possibly caused by $B_0$ inhomogeneity. Such maps are complimentary but critical to quantification and interpretation of the separated mono- and bi-$T_2$ sodium signals.

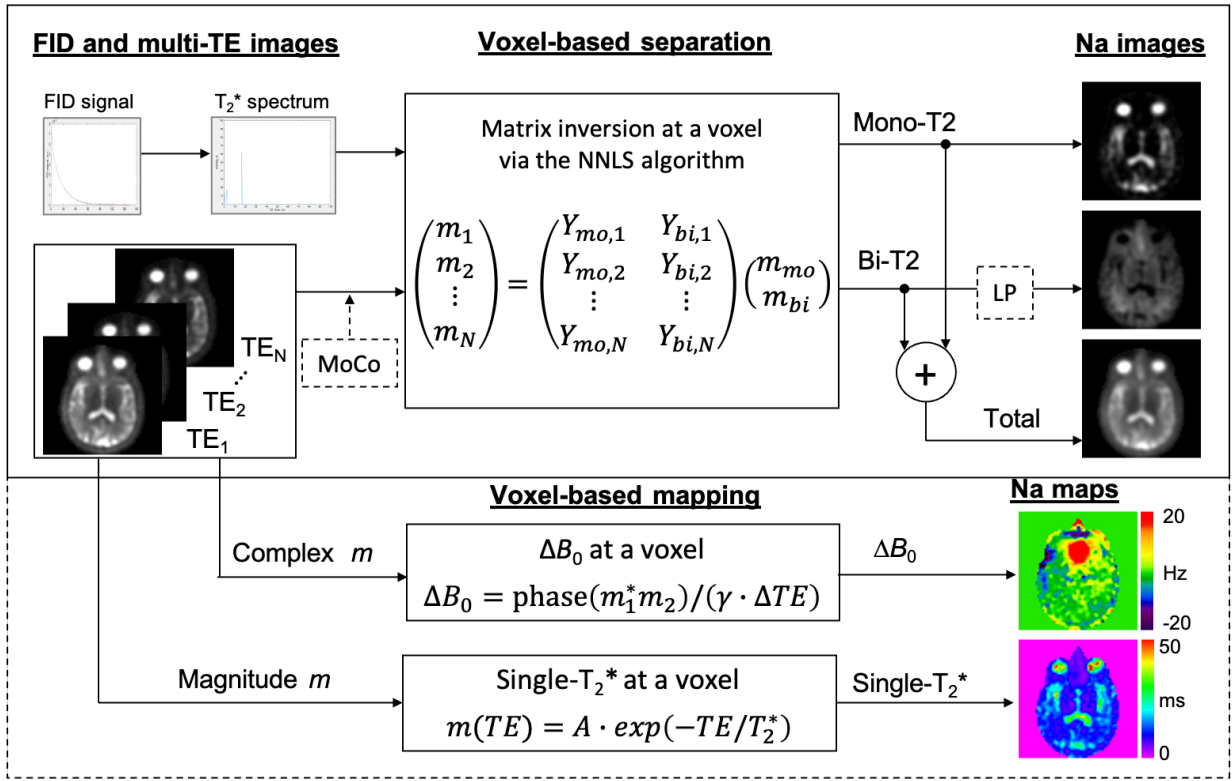

**Fig. 1. Flowchart of the proposed MSQ sodium MRI. Input:** multi-TE SQ images $m(TE)$ and an FID signal producing $T_2^*$ spectrum. Motion correction (MoCo) between SQ images is optional. **Separation:** matrix inversion voxel-by-voxel across field-of-view (FOV). **Output:** three sodium images (mono-$T_2$ as $m_{mo}$, bi-$T_2$ as $m_{bi}$, and total as $m_{mo}+m_{bi}$). The low-pass (LP) filtering (optional) is a 3D smoothing of size 3×3×3 or others, to further reduce random noise. **Additional outputs:** the maps of $B_0$ inhomogeneity and single-$T_2^*$.

*Human studies*

To demonstrate feasibility of the proposed MSQ approach at clinical field strength of 3 Tesla, we conducted the human study on 15 subjects including nine healthy adults (age 39.6±21.4 years ranging 21–74 years; 3 males and 6 females) and six patients with diverse neurological conditions (1 bipolar disorder, 3 epilepsy, 1 multiple sclerosis, and 1 mild traumatic brain injury; age 30.5±15.1 years ranging 18–59 years; 3 males and 3 females), after the exclusion of one healthy subject and one patient due to motion between the two TE-images. The MSQ sodium MRI was performed on a clinical 3T MRI scanner (Prisma, Siemens Healthineers, Erlangen, Germany) with a custom-built 8-channel dual-tuned ($^1$H-$^{23}$Na) head array coil (*32*).

Fig. 2 presents a typical case from a young healthy female of age 26 years and includes 3D sodium images of the brain at $TE_1/TE_2=0.3/5$ms (Fig. 2A), FID signal of whole brain and associated fitting error and $T_2^*$ spectrum (Fig. 2B), the separated sodium images from the two-TE images using $(T_{2,mo}^*, T_{2,bs}^*, T_{2,bl}^*) = (50.0, 6.0, 19.0)$ ms (Fig. 2C), and inverse-contrast displays (Fig. 2D). In Figs. 2E-G are SNR, $\Delta B_0$, and single-$T_2^*$ maps calculated from the two-TE images in Fig. 2A. Fig. 2 indicates that signals from CSF in the brain were effectively separated into mono-$T_2$ sodium image (Fig. 2C or 2D), while signals from brain tissues such as gray and white matters were separated into bi-$T_2$ sodium image (Fig. 2C or 2D). Notably and surprisingly, signal intensity across brain tissues looks more uniform in the bi-$T_2$ sodium images than in the mono-$T_2$ sodium images (Fig. 2C), total sodium images (Fig. 2C), and $TE_1$-images (Fig. 2A). SNR in Fig. 2E is ≥25 in most regions of the brain, ensuring a robust separation as suggested in the simulations in Extended Data (Fig. E4). The field inhomogeneity $\Delta B_0$ in Fig. 2F varied between ±20 Hz across the brain, with the largest off-resonance in the prefrontal and occipital lobes, leading to visible blurring of the tissues in the bi-$T_2$ sodium images (Fig. 2C or 2D, sagittal). The single-$T_2^*$ map in Fig. 2G provides a spatial distribution of short and long $T_2^*$ components across the brain, complementary to $T_2^*$ spectrum in Fig. 2B. It also indicates that majority of long $T_2^*$ components are located in the prefrontal lobe in this particular case (Fig. 2G, sagittal).

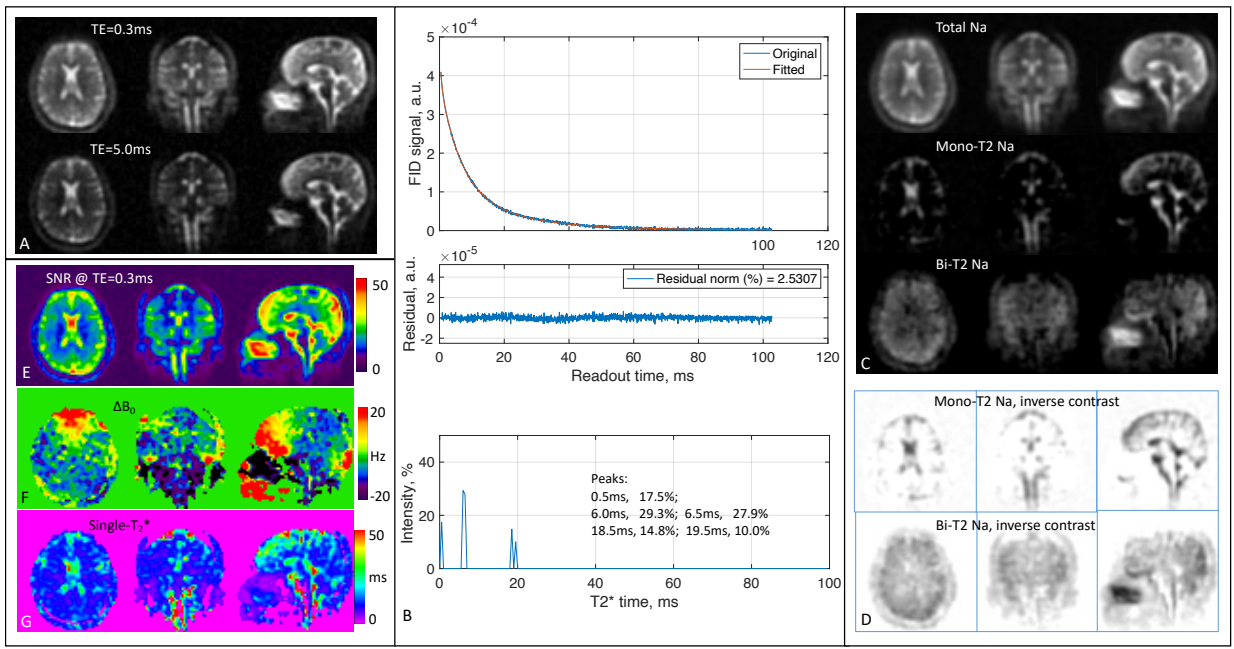

**Fig. 2. Human study #1 (26-year-old female, healthy). (A)** 3D sodium images of the brain in three orthogonal slices at $TE_1/TE_2=0.3/5$ms. **(B)** FID signal of whole brain and associated fitting error and $T_2^*$ spectrum. **(C)** Separated sodium images from the 2-TE images in A using $(T_2^*_{mo}, T_2^*_{bs}, T_2^*_{bl}) = (50.0, 6.0, 19.0)$ ms according to peaks in B. **(D)** Inverse-contrast display to highlight areas of low intensity. All the images in A and C were displayed using the same window/level. **(E-G)** Maps of SNR, $\Delta B_0$, and single-$T_2^*$, calculated from the 2-TE images in A.

Fig. 3 shows potential benefits from bi-$T_2$ sodium images of patients with neurological disorders such as bipolar disorder which is known to cause abnormally-high intracellular sodium concentration in the brain but locations are hard to define (*33, 34*). This case reports a middle-aged male patient of age 59 years with self-reported bipolar disorder. The separated sodium images were attained at $(T_{2,mo}^*, T_{2,bs}^*, T_{2,bl}^*)$ = (50.0, 2.5, 7.0) ms according to the peaks in Fig. 3B. The bi-$T_2$ sodium images (Fig. 3C or 3D) clearly highlighted brain regions of an elevated bi-$T_2$ sodium signal against surrounding tissues, with a ratio of 1.78 vs. 1.40 (or 27.1% increase) before the separation (Fig. 3C). These regions have no visible contrast in total or $TE_1$-images (Fig. 3A or 3C). SNR in these regions is >40 (Fig. 3E), supporting a robust separation. The field inhomogeneity $\Delta B_0$ in these regions is low (<5 Hz, Fig. 3F), excluding field-induced artifacts. Single-$T_2^*$ map in Fig. 3G shows abnormally low $T_2^*$ values in the regions, confirming an increase in short-$T_2^*$ components.

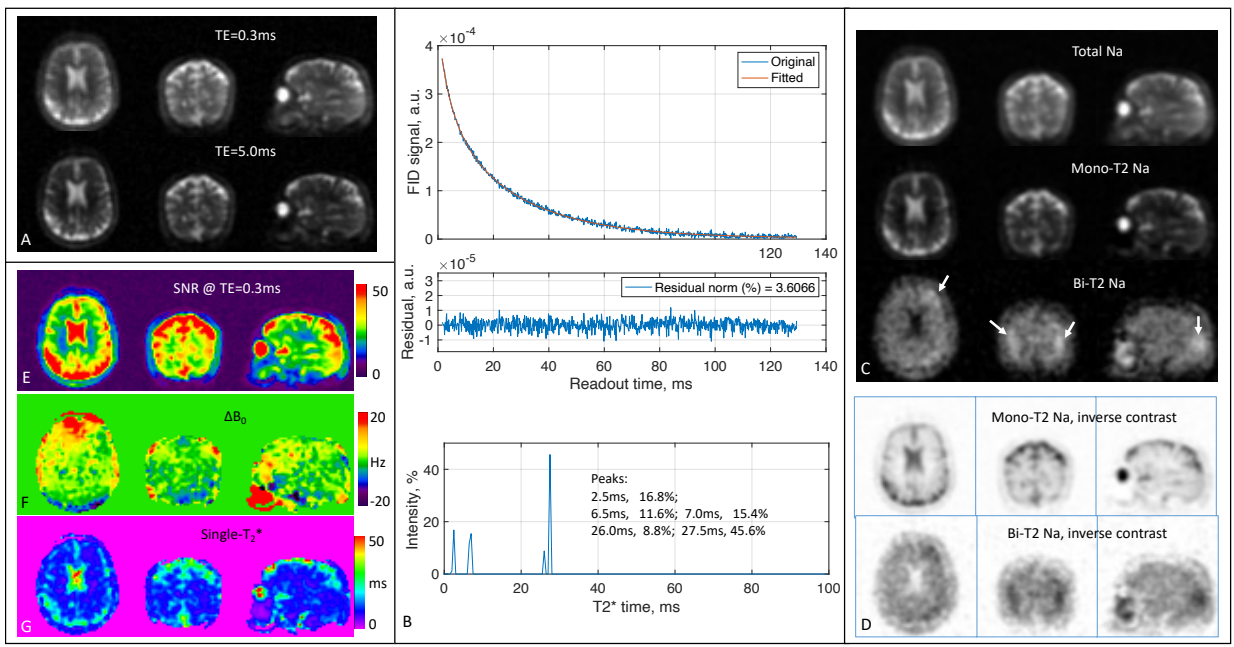

**Fig. 3. Human study #2 (59-year-old male, bipolar disorder patient).** (**A**) 3D sodium images of the brain at $TE_1/TE_2$=0.3/5ms. (**B**) FID signal of whole brain and associated fitting error and $T_2^*$ spectrum. (**C**) Separated sodium images from the 2-TE images in A using $(T_2^*{}_{mo}, T_2^*{}_{bs}, T_2^*{}_{bl})$ = (50.0, 2.5, 7.0) ms according to peaks in B. (**D**) Inverse-contrast display to highlight areas of low intensity. All the images in A and C were displayed using the same window/level, except bi-$T_2$ sodium where W/L was halved. (**E-G**) Maps of SNR, $\Delta B_0$, and single-$T_2^*$, calculated from the 2-TE images in A. Note that the bi-$T_2$ sodium images in C highlight possible regions (arrows) of elevated bi-$T_2$ sodium concentration in the parenchyma.

**Discussion** (510 words)

We presented here a new technique MSQ to separate mono- and bi-$T_2$ sodium signals with high accuracy offered by matrix inversion voxel by voxel (Fig. 1). The MSQ approach leverages intrinsic difference in $T_2$ relaxation between the two sodium populations (Figs. 2, 3). The 2-TE sampling scheme stands out for smaller noise transfer during the separation (Extended Data Fig. E3). The $T_2$* spectrum of whole brain has sparse peaks and confirms a global set of $T_2$* values $\{T_{2,mo}^*, T_{2,bs}^*, T_{2,bl}^*\}$ applicable to humans (Extended Data Fig. E1). The measurement of $T_2$* spectrum facilities fine tuning of the global set of $T_2$* values towards individual subjects of diverse $T_2$* relaxations in the brain (Figs. 2, 3).

However, a global set of $T_2$* values may not be plausible in such situations where $T_2$* in an individual brain has substantial spatial variation (*37–44*), multi-regional sets, or a linear combination of them, may be applied to the separation.

Of note, subsequent quantification of sodium concentrations after the separation is not addressed in this study because such quantification involves complicated procedures including calibration to sodium concentration from signal and corrections for inhomogeneous coil sensitivity. These procedures deserve a separate study to adequately address in detail.

The MSQ approach has limitations and understanding them is crucial to practice of the approach. The first limitation is the two-population model (mono- and bi-$T_2$) which is of risk to produce a false-positive error for bi-$T_2$ sodium. For instance, if there is no bi-$T_2$ but rather two mono-$T_2$ sodium decays of different $T_2$* values within a voxel, they would be falsely separated into bi-$T_2$ sodium because they can be combined mathematically into a bi-exponential decay. This kind of false positive error stems from the fact that the separation is based on *mathematical* model, instead of *physical* model as does the TQF separation.

The second limitation may occur at voxels filled solely with mono-$T_2$ sodium of very small $T_2$* value, such as regions in nose and sinuses (Figs. 2C and 3C). Such areas may mislead the MSQ to produce pseudo bi-$T_2$ sodium. These mis-separated regions can be readily identified by cross-referencing the maps of $\Delta B_0$ and single-$T_2$* (Figs. 2 and 3).

The third limitation is the underestimate of bi-$T_2$ sodium signal caused by the image at $TE_1$, as illustrated in the phantom studies (Extended Data Fig. E5). The separation Eq. (2) assumes $TE_1$-image intensity precisely at $TE_1$ (i.e., a very short readout time). Actual $TE_1$-image intensity is an average over readout time during which short-$T_2$* components decay significantly when readout is relatively long, such as readout time $T_s$=36.32ms about ten times long of a short-$T_2$* of 3ms seen in this study. Therefore, the extent of underestimation depends on readout time or pulse sequence.

To mitigate this problem, two strategies may be applicable. One is to replace $TE_1$ in Eq. (2) with an effective (larger) value accounting for short-$T_2^*$ decay during readout. The other is to shift $T_{2,bs}^*$ to a larger value, as did in previous work on phantoms (*30*). Alternatively, correction for such underestimation may be integrated into calibration process when calculating sodium concentration (Extended Data Fig. E5).

**Materials and methods** (955 words)

*Experimental Design*

The objective of this study is to demonstrate the feasibly of proposed MSQ approach to separation of mono- and bi-$T_2$ sodium signals from human brains at a clinical field strength of 3 Tesla. We designed three types of experiments: numerical simulations, phantom experiments, and human subject studies. To implement the MSQ approach, we used a clinical MRI scanner at 3 Tesla for data acquisition and a custom-developed software package for data processing.

*Clinical MRI scanner for data acquisition*

The sodium MRI images and FID signals were acquired on two 3T clinical MRI scanners for the phantom experiments and human studies, respectively. One was MAGNETOM Trio Tim (Siemens Medical Solutions, Erlangen, Germany) with a dual-tuned ($^1$H-$^{23}$Na) volume head coil (Advanced Imaging Research, Cleveland, OH), and used for the phantom experiments. The other was MAGNETON Prisma (Siemens Healthineers, Erlangen, Germany) with a custom-built 8-channel dual-tuned ($^1$H-$^{23}$Na) head array coil (*32*), and used for human studies.

*Software for data processing and image display*

We developed a software (code) called *SepMoBi* in MATLAB (R2021a, MathWorks, Natick, MA) on a laptop computer (MacBook Pro, 16GB memory, Apple M1 chip, Apple Inc., Cupertino, CA), and used it for data processing, including the calculations of $T_2^*$ spectrum, mono- and bi-$T_2$ sodium images, and maps of SNR, $\Delta B_0$ and single-$T_2^*$. In addition, we used software *MRView* (MRI Research, Mayo Clinic and Foundation, December, 2020) for image display and parameter calculation in ROIs (Extended Data Fig. E6).

*Pre-processing of FID signals*

When acquired with an array coil, FID signals may have unique initial phases $\{\varphi_{0,l}, l=1, 2, …, N_c\}$ at individual elements, and need to be aligned to produce a resultant FID signal. Alignment (via phase correction) can be towards a reference phase such as zero phase, one of the initial phases, or mean phase across elements. In addition, signal intensity at individual elements needs to be scaled using "FFT factor" stored in the header of a raw FID data file, for instance.

FID signals at the first few samples are distorted by hardware filtering during analog-to-digital conversion (ADC). The number of affected samples are in a range of 3–10 points, dependent on sampling bandwidth, with the first sample having the largest distortion. This distortion alters measurement of $T_2^*$ components, especially the short $T_2^*$ components which are critically important to the bi-$T_2$ sodium. Correction for the distortion can be performed using an established exponential extrapolation (see Supporting Materials).

*Global $T_2^*$ spectrum*

We measured FID signals on clinical 3T MRI scanners using a product pulse sequence, either *AdjXFre* embedded in manual shimming or independent *fid_23Na*, with acquisition parameters: TE=0.35–1.0ms, TR=100–300ms, and averages =1–128, and TA=0.2–39s. After the pre-processing described above, resultant FID signals are curve-fitted to Eq. (4) using the NNLS algorithm (*27*) when $T_2^*$ values are pre-distributed in a range of interest [$T_2^*{}_{min}$, $T_2^*{}_{max}$] at uniform or non-uniform intervals $\{\Delta T_{2,j}^*, j = 1, 2, ...\}$. Amplitudes $\{A_j\}$, called $T_2^*$ spectrum, determine relative incidence of $T_2^*$ components in an imaging volume which counts all $T_2^*$ components from both mono- and bi-$T_2$ sodium populations. We used a uniform interval of $\Delta T_2^* = 0.5$ms in a range of interest 0.5–100 ms for a high resolution of $T_2^*$ values. Assignment of the peaks in the $T_2^*$ spectrum to $\{T_{2,mo}^*, T_{2,bs}^*, T_{2,bl}^*\}$ was based on their relative positions of $T_{2,mo}^* \geq T_{2,bl}^* \gg T_{2,bs}^*$ and intensity ratio about 6:4 for the bi-$T_2$ sodium. Robustness of the $T_2^*$ measurement including data acquisition and spectrum computation is addressed in Supporting Information.

*Human studies*

The human studies were approved by local Institutional Review Board (IRB) at NYU Grossman School of Medicine, New York, NY, USA, and performed in accordance with relevant guidelines and regulations. Informed consent was obtained from all subjects. For data acquisition, the same TPI pulse sequence was used as in the phantom studies. Sodium images were reconstructed off-line using the gridding algorithm (*47, 48*), channel by channel, and combined into a resultant image via the sum-of-squares (SOS) algorithm (*49*). To decouple random noise across channels, an orthogonal linear transform (detailed in Ref. *46*) was performed in which physical channel data were transformed into virtual channels with random noise independent from channel to channel. This decoupling and denoising process also normalized signal amplitudes across channels by dividing noise standard deviation. Separation of the mono- and bi-$T_2$ sodium signals was then implemented in the same way as in the phantom studies.

*Mapping of $\Delta B_0$ and single-$T_2^*$*

To map $\Delta B_0$ or $\Delta f_0$ (=$\gamma \Delta B_0/2\pi$), Hermitian product method (*50*) was performed via Eqs. (10–12) at individual imaging voxels to calculate phase differences $\{\Delta\varphi_i, i = 1, 2, \ldots, N-1\}$ between TEs $\{TE_i, i = 1, 2, \ldots, N\}$. Image amplitude at individual channels were corrected with the scale factors $\{w_l, l = 1, 2, \ldots, N_c\}$. Phase unwrapping was not performed due to small intervals in the TEs and, in general, small inhomogeneity in $B_0$ field in sodium MRI. Computation for $\Delta B_0$ map is fast (0.078s) on a Mac laptop computer for images of matrix size 64×64×64 at two TEs.

$$\Delta f_0 = \frac{1}{2\pi(N-1)} \sum_{i=1}^{N-1} \Delta\varphi_i / \Delta TE_i \qquad (10)$$

$$\Delta\varphi_i = phase\{\sum_{l=1}^{Nc} w_l^2 \cdot m_l^*(TE_i) \cdot m_l(TE_{i+1})\} \qquad (11)$$

$$\Delta TE_i = TE_{i+1} - TE_i \qquad (12)$$

To map single-$T_2^*$, a MATLAB curve-fitting function *fit(x, y, 'exp1')* was used to calculate single-$T_2^*$ values at each voxel via Eq. (13). A restriction ($T_2^*{}_{max}$ < 100ms) was enforced to exclude unreasonable values caused by random noise. The computation time is acceptable (10min17s).

$$|m(TE_i)| = A_0 \exp(-TE_i/T_2^*), \quad 0 \leq T_2^* \leq T_{2,max}^* \qquad (13)$$

*Calculation of signal-to-noise ratio (SNR)*

In a region of interest (ROI), SNR was calculated via Eq. (14) in a simplified way for both volume and array coils by taking the ratio of mean intensity *S* in a ROI to noise standard deviation (SD) in noise-only background regions. A factor of 0.655 was applied to noise SD to account for Rician distribution in magnitude images (*51*). For SNR mapping, pixel signal is used in the calculation.

$$SNR = 0.655\ S/SD \qquad (14)$$

*Statistical Analysis*

A regular statistical significance ($P$=0.05) was applied to the comparisons, via Student's *t*-test, between the two sets of data in this work. Minimum sample size for the *t*-test is 16, with 80% power, 5% significance level, two-sided test, and 1.0 effect size (*29*).

# Extended Data for

# Multi-TE Single-Quantum Sodium ($^{23}$Na) MRI: A Clinically Translatable Technique for Separation of Mono- and Bi-T$_2$ Sodium Signals

Yongxian Qian* *et al.*
.

*Corresponding author. Email: Yongxian.Qian@nyulangone.org

**This PDF file includes:**

    Extended text, figures, and tables
    Text
    Figs. E1 to E6
    Table (no)
    References (no)

## $T_2$* values in whole brain of the study subjects

The MSQ approach needs an input of $T_2$* values for mono-$T_2$ sodium, $T^*_{2,mo}$, and for bi-$T_2$ sodium, $T^*_{2,bs}$ and $T^*_{2,bl}$, at *each* voxel $\Delta V$. Measurement of these values however is time consuming (~2 hours) and impractical in clinic routine. Alternatively, we use a global set of $T_2$* values $\{T^*_{2,mo}, T^*_{2,bs}, T^*_{2,bl}\}$ for *all* voxels. This global set can be measured quickly (<1 min) on whole brain of each study subject by acquiring free induction decay (FID) signal $s(t)$ and fitting it with multi-term exponential $T_2$* decays to attain a $T_2$* spectrum through Eq. (E1).

$$|s(t)| = \sum_j A_j \exp(-t/T^*_{2,j}) \tag{E1}$$

Two representative FID signals and $T_2$* spectra are illustrated in Fig. E1A for a healthy young subject (21 years old, male) and Fig. E1B for a patient with epilepsy (31 years old, male). The spectra are sparse with just 2–4 peaks, indicating that the global set of $\{T^*_{2,mo}, T^*_{2,bs}, T^*_{2,bl}\}$ is a reasonable estimate for the separation. Notably, these $T_2$* values are different from subject to subject as shown in the two-dimensional (2D) scatter plot (Fig. E1C) from all 15 study subjects including nine healthy subjects and six patients. The short-$T_2$* component is clearly crowded in a range of 1–5 ms while the long-$T_2$* is widely scattered in three bands centered at 10, 20, and 30ms, respectively. Interestingly, the long-$T_2$* component is shifted to lower values in the patient group, relative to the healthy group. There is no difference between males and females in the healthy group. Therefore, $T_2$* values are heterogeneous across subjects and measurement of $T_2$* spectrum is necessary for each subject.

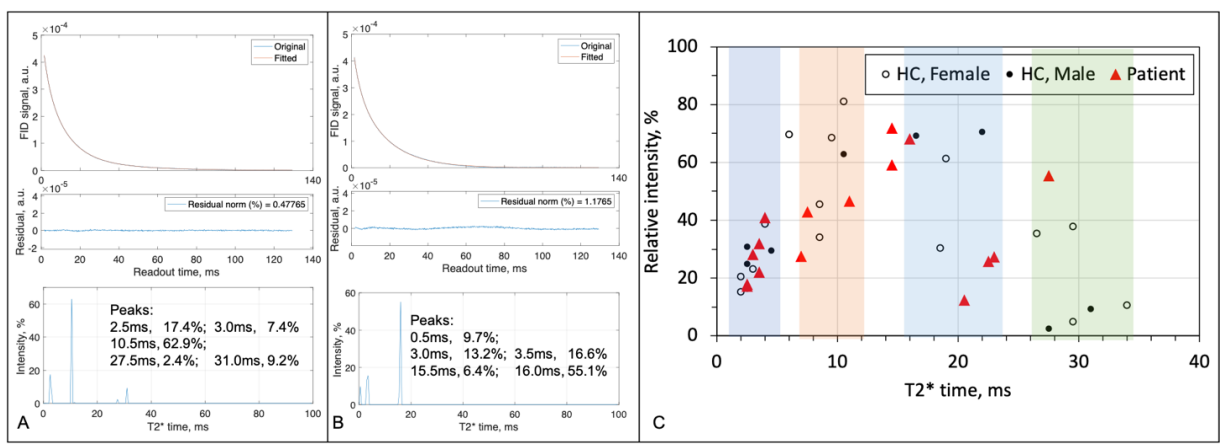

**Fig. E1. Human study: 2D scatter plot of individual $T_2$* components across all study subjects. (A)** A typical $T_2$* spectrum and associated FID signal and fitting error from whole brain of a healthy young subject (21 years old, male). **(B)** Another example from a patient with epilepsy (31 years old, male). **(C)** Scatter plot of individual $T_2$* components.

## Sensitivity of the separation to $T_2^*$ values

What would happen for the separation if a global set of $T_2^*$ values measured on whole brain is not the same as local one at a voxel? How sensitive the separation is to $T_2^*$ values? Intuitively, the solution $\{m_{mo}, m_{bi}\}$ to Eq. (1) would not be very sensitive to $T_2^*$ values due to exponential decays in Eqs. (1A, 1B). This observation can be verified theoretically and numerically.

Theoretically, small changes in $T_2^*$ values, $\{\delta T_{2,q}^*, q=mo, bs, bl\}$, lead to small changes $\{\delta m_p, p=mo, bi\}$ in $m_{mo}$ and $m_{bi}$ under the same $m(t)$, that is,

$$0 = \delta(m_{mo}Y_{mo}) + \delta(m_{bi}Y_{bi}) \tag{E2}$$

$$-m_\delta = Y_{mo}\delta m_{mo} + Y_{bi}\delta m_{bi} \tag{E2A}$$

$$m_\delta \equiv m_{mo}\delta Y_{mo} + m_{bi}\delta Y_{bi} \tag{E2B}$$

$$\delta Y_q = \left(\frac{tY_q}{T_{2,q}^*}\right)\left(\frac{\delta T_{2,q}^*}{T_{2,q}^*}\right) \leq e^{-1}\left(\frac{\delta T_{2,q}^*}{T_{2,q}^*}\right), \quad q=mo, bs, bl \tag{E2C}$$

$$\delta Y_{bi} = a_{bs}\,\delta Y_{bs} + a_{bl}\,\delta Y_{bl} \leq e^{-1}\left[a_{bs}\left(\frac{\delta T_{2,bs}^*}{T_{2,bs}^*}\right) + a_{bl}\left(\frac{\delta T_{2,bl}^*}{T_{2,bl}^*}\right)\right] \tag{E2D}$$

Where $\delta$ is difference operator.

Numerically, errors in $m_{mo}$ and $m_{bi}$ can be calculated, given a series of $\{T_{2,q}^*, \delta T_{2,q}^*; q=mo, bs, bl\}$ at a specific pair $(m_{mo}, m_{bi})$. This creates a plot showing how $\{m_{mo}, m_{bi}\}$ changes with $\{\delta T_{2,q}^*; q=mo, bs, bl\}$.

Fig. E2 shows two extreme cases: mono-$T_2$ sodium dominating at $m_{mo} = 0.9$ and bi-$T_2$ sodium dominating at $m_{bi} = 0.9$. In Fig. E2A, an error in $T_{2,mo}^*$ causes an error in $m_{mo}$ or $m_{bi}$ much smaller for the dominant one, e.g., $\delta m_{bi} < 2.2\%$ when $\delta T_{2,mo}^* < 20\%$, and $\delta m_{mo} < 2.9\%$ when $\delta T_{2,mo}^* < 5.0\%$. In Fig. E2B, an error in $T_{2,bs}^*$ has a small impact on both $m_{mo}$ and $m_{bi}$, e.g., when dominating, $\delta m_{bi} < 4.8\%$ and $\delta m_{mo} < 0.04\%$ when $\delta T_{2,bs}^* < 20\%$. In Fig. E2C, an error in $T_{2,bl}^*$ leads to an error in $m_{mo}$ or $m_{bi}$ much smaller for the dominant one, e.g., $\delta m_{bi} < 5.2\%$ and $\delta m_{mo} < 0.6\%$ when $\delta T_{2,bl}^* < 20\%$. The best case is Fig. E2B where $T_{2,bs}^*$ has small impact (<4.9%) on both mono- and bi-$T_2$ sodium signals. The worst case is Fig. E2A (top) where $T_{2,mo}^*$ had a large impact on bi-$T_2$ sodium signal, $\delta m_{bi} = 35.6\%$ when $\delta T_{2,mo}^* = -5\%$. In other words, when the mono-$T_2$ sodium is very dominating, $T_{2,mo}^*$ value should be as accurate as possible (achievable via single-$T_2^*$ map) to attain the best separation for the bi-$T_2$ sodium.

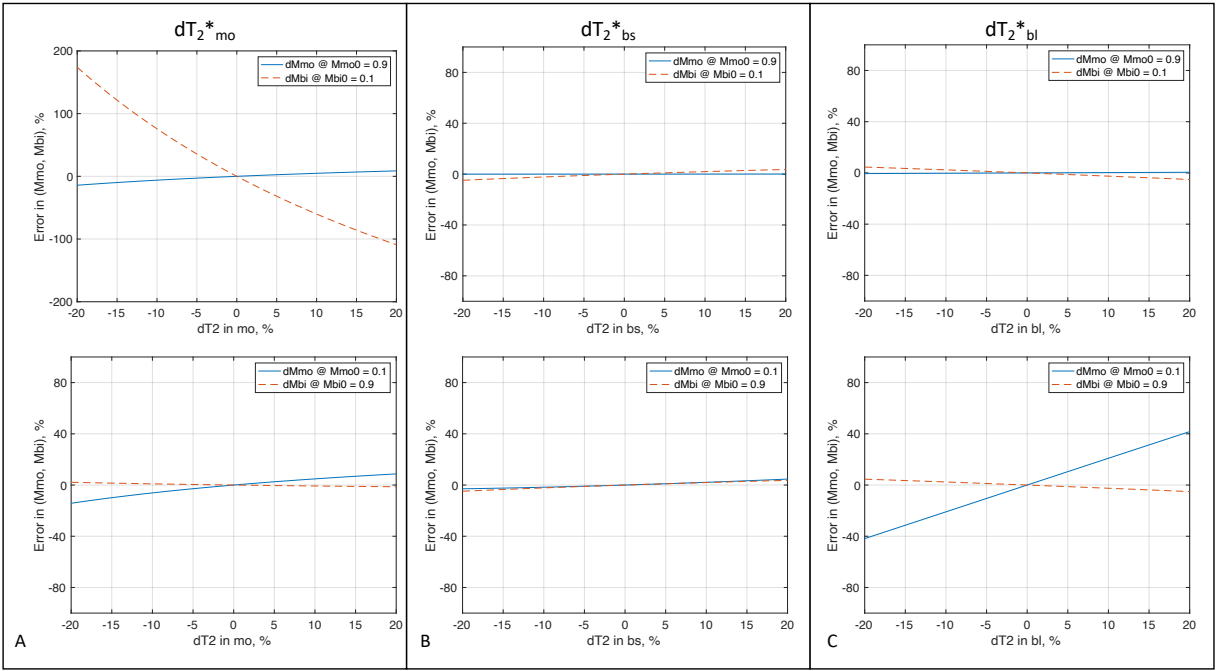

**Fig. E2. Simulated impact of $T_2^*$ values on separation of mono- and bi-$T_2$ sodium signals.** The impact of individual $T_2^*$ components on the separation of sodium signals ($m_{mo}$, $m_{bi}$) at $\{T_2^*_{mo}, T_2^*_{bs}, T_2^*_{bl}\}$ = (50.0, 3.5, 15.0) ms for two extreme cases: mono-$T_2$ sodium dominating (top row), $m_{mo}$=0.9, and bi-$T_2$ sodium dominating (bottom row), $m_{bi}$=0.9. **(A)** An error in $T_2^*_{mo}$ produces an error in $m_{mo}$ or $m_{bi}$ much smaller for the dominant signal. **(B)** An error in $T_2^*_{bs}$ has a small impact on both $m_{mo}$ and $m_{bi}$. **(C)** An error in $T_2^*_{bl}$ leads to an error in $m_{mo}$ or $m_{bi}$ much smaller for the dominant one. The best case is Column B where the $T_2^*_{bs}$ had small impact (<4.9%) on both mono-$T_2$ and bi-$T_2$ sodium signals. The worst case is (A) where the $T_2^*_{mo}$ had a large impact on the bi-$T_2$ sodium signal, $\Delta m_{bi}$ = 35.6% when $\Delta T_2^*_{mo}$ = -5%.

## *Optimization of the number of TEs*

In principle, the more TEs the better differentiation between $T_2^*$ relaxations of the mono- and bi-$T_2$ sodium populations. In practice, the number of TEs is restricted by total scan time (TA), SNR, signal decay, and risk of motion across TEs. Therefore, a trade-off must be made for the number of TEs. To determine an optimal number of TEs, it is necessary to understand noise propagation in Eq. (2). We applied singular value decomposition (SVD) analysis (*28*) on matrix **Y** in Eq. (2), i.e.,

$$\mathbf{Y} = \mathbf{U}\,\mathbf{\Sigma}\,\mathbf{V}^T \tag{E3}$$

$$\mathbf{\Sigma} = \text{diag}(\sigma_1, \sigma_2) \tag{E3A}$$

$$\mathbf{X} = \mathbf{V}\,\mathbf{\Sigma}^{-1}\,\mathbf{U}^T\,\mathbf{M} \tag{E3B}$$

Singular values ($\sigma_1 \geq \sigma_2 \geq 0$) determine noise transfer (amplification or suppression) in Eq. (E3C) from the measured TE-images **M** to the separated mono- and bi-$T_2$ sodium images **X**. However, Eq. (E3B) allows negative values in **X** when random noise contaminates **M**. This violates the "non-negative" condition on **X**. Therefore, the SVD analysis is applicable only to **X** elements with SNR $\geq 2$ where the elements, with Gaussian noise, have 95.4% of chance in the territory of non-negative value (*29*).

Fig. E3 presents the results at a typical set of $T_2^*$ values, $\{T_{2,mo}^*, T_{2,bs}^*, T_{2,bl}^*\} = \{50.0, 3.5, 15.0\}$ms for three cases of interest: an ideal case of 80 TEs, practical case #1 of 8 TEs, and practical case #2 of two TEs. Singular value $\sigma_2$ is less than 1.0 for the 8-TE and 2-TE cases, leading to an amplification of noise. Therefore, a better choice for less noise amplification is the 2-TE case, in which $TE_2$ at 5ms produced a value near maximum of $\sigma_2$ while preserving higher signal than larger $TE_2$. Thus, the 2-TE scheme was selected for human studies.

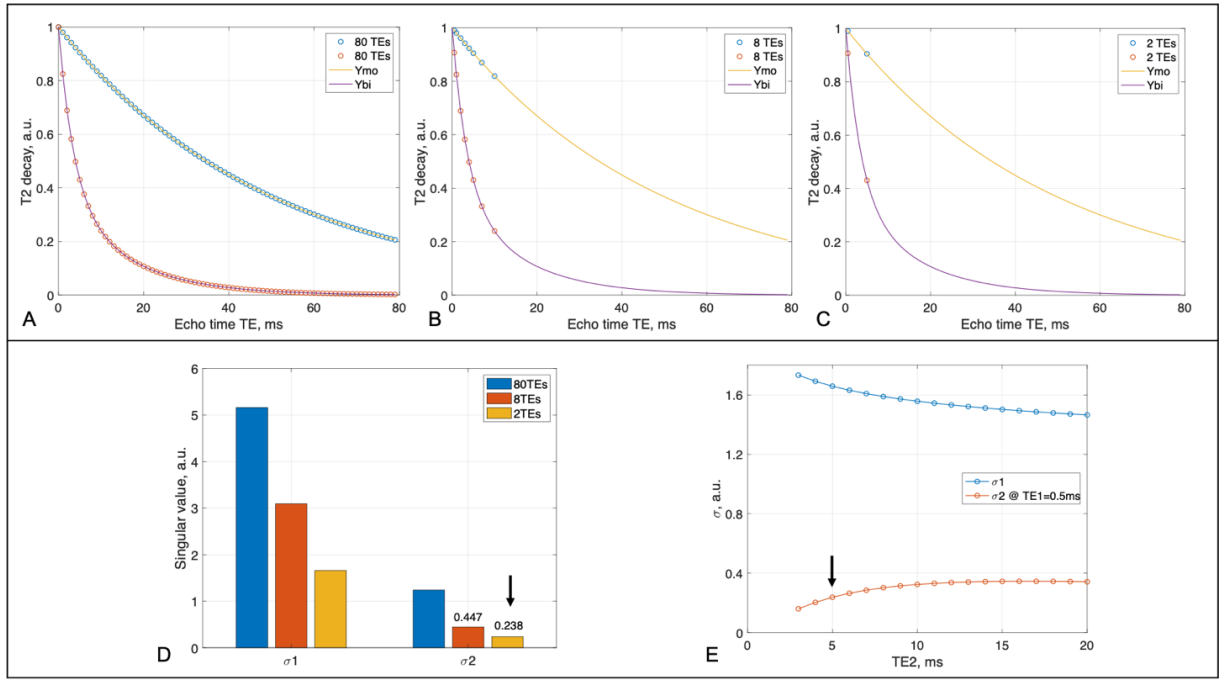

**Fig. E3. Optimization of TE sampling scheme via SVD analysis**. **(A)** A reference sampling scheme of 80 TEs in a range of 0–79 ms at an interval of 1.0ms, distributing on a mono-exponential $T_2^*$ decay $Y_{mo}$ (TE) of mono-$T_2$ sodium at a typical value of $T_2^*{}_{mo}$=50ms, and on a bi-exponential $T_2^*$ decay $Y_{bi}$ (TE) of bi-$T_2$ sodium at a typical set $\{T_2^*{}_{bs}, T_2^*{}_{bl}\} = \{3.5, 15.0\}$ms. **(B)** An intuitively-favorable scheme of 8 TEs at $\{0.5, 1, 2, 3, 4, 5, 7, 10\}$ms, distributing on the decay curves $Y_{mo}$ and $Y_{bi}$. **(C)** The optimal scheme of 2 TEs at $\{0.5, 5\}$ms. **(D)** SVD singular values ($\sigma_1$ and $\sigma_2$) of the three TE sampling schemes. **(E)** Singular values of the 2-TE scheme changing with the 2nd TE (or TE$_2$). In D, $\sigma_2$ is less than 1.0 at the 8-TEs and 2-TEs, leading to noise amplification. Thus, a better choice for less noise amplification is the 2-TEs (arrow). In E, TE$_2$ at 5ms (arrow) produced a value near maximum for $\sigma_2$ while preserving higher signal than the larger TE$_2$. Therefore, the 2-TEs scheme is an optimal one for the human brain.

*Numerical simulations*

The separation of mono- and bi-$T_2$ sodium signals was first investigated on numerical models via Eq. (3) at a typical set of $T_2^*$ values, $(T_{2,mo}^*, T_{2,bs}^*, T_{2,bl}^*) = (50.0, 3.5, 15.0)$ ms, and an optimal two-TE scheme, TE = (0.5, 5.0) ms. Sodium image signals were calculated via Eq. (1), with an additive Gaussian noise, $N(0, \sigma^2)$, at each of noise trials (independent from each other), $m(t) + n(t)$. The mono- and bi-$T_2$ signal intensities $\{m_{mo}, m_{bi}\}$ vary in a normalized range of 0.1–0.9 at a step size = 0.1. The separation was implemented using the function *lsqnonneg*() in MATLAB, and repeated $N_{noise}$ times at each of the specific intensity. Mean and standard deviation (SD) were reported as the separated sodium signal. $N_{noise}$=1054 was chosen to detect a 10% of SD, or 0.1 effect size d=Δμ/SD, in difference between the mean and true value at 90% power and 5% significant level under the two-sided Student's *t*-test (*29*).

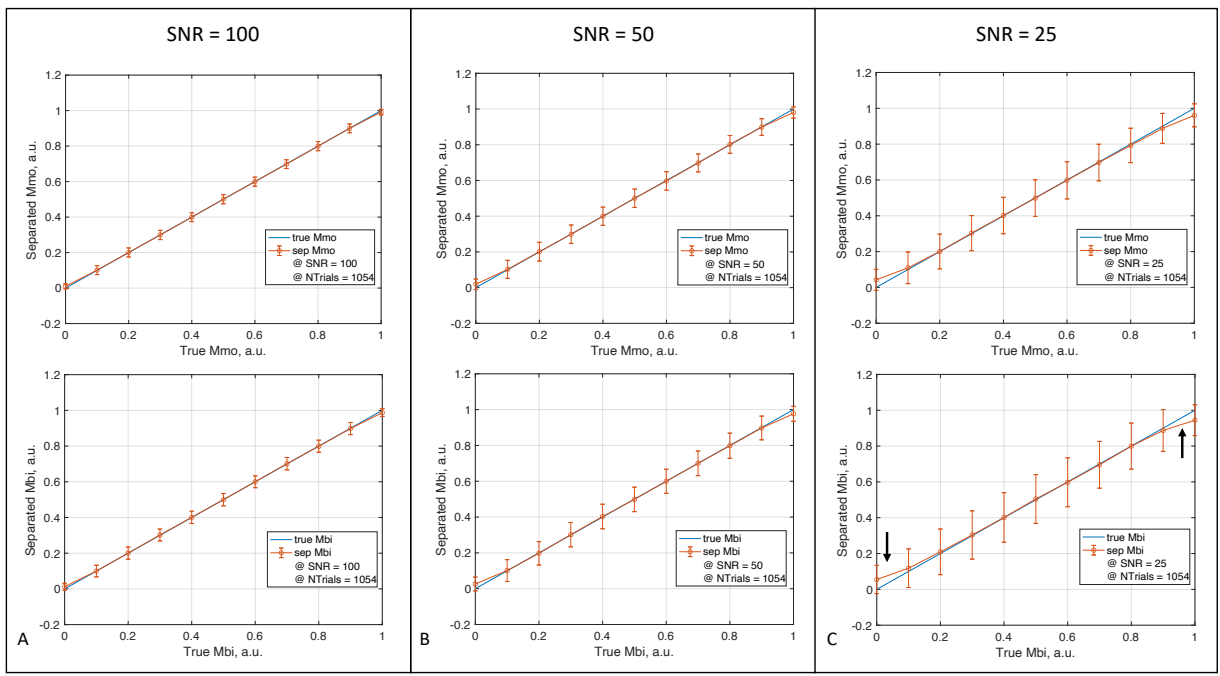

**Fig. E4. Simulated separation of the mono- and bi-$T_2$ sodium signals.** Sodium signals ($m_{mo}$, $m_{bi}$) were simulated at a typical set of $(T_2^*_{mo}, T_2^*_{bs}, T_2^*_{bl}) = (50.0, 3.5, 15.0)$ ms and the 2-TE scheme TE=(0.5, 5.0). **(A)** Extra-high SNR = 100. **(B)** High SNR = 50. **(C)** Regular SNR = 25. The standard deviation (error bar) of the separated $m_{mo}$ (top row) and $m_{bi}$ (bottom row) consistently decreased with SNR increasing from 25 to 100. There was an underestimate (3.9–5.6%, arrow) for $m_{mo}$ or $m_{bi}$ near maximum value 1.0, but an overestimate (4.2–5.5%, arrow) near minimum value 0.0, with an amount decreasing with SNR increasing.

Fig. E4 demonstrates the simulated separation at three SNRs of interest: 100 (extra-high), 50 (high), and 25 (regular). The SD (error bar) consistently decreased with SNR increasing. There was an underestimate in $m_{mo}$ and $m_{bi}$ near maximum value 1.0, but an overestimate near minimum value 0.0 with an amount decreasing with SNR increasing.

*Phantom studies*

The separation was then investigated on four phantoms with known sodium concentrations, which were custom-built and described in previous work (*25, 30*). The phantoms are 50-mL centrifuge tubes filled with a mixture of distilled water, 10% w/w agar powder, and sodium chloride (NaCl) at three concentrations of 90, 120, and 150 mM and at 150mM without agar, to mimic bi- and mono-$T_2$ sodium signals in human brains. Sodium MRI was performed on a clinical scanner at 3T (MAGNETOM Trio Tim, Siemens Medical Solutions, Erlangen, Germany) with a dual-tuned ($^1$H-$^{23}$Na) volume head coil (Advanced Imaging Research, Cleveland, OH). The data acquisition was implemented using an SNR-efficient, three-dimensional (3D) pulse sequence called twisted projection imaging (TPI) (*31*).

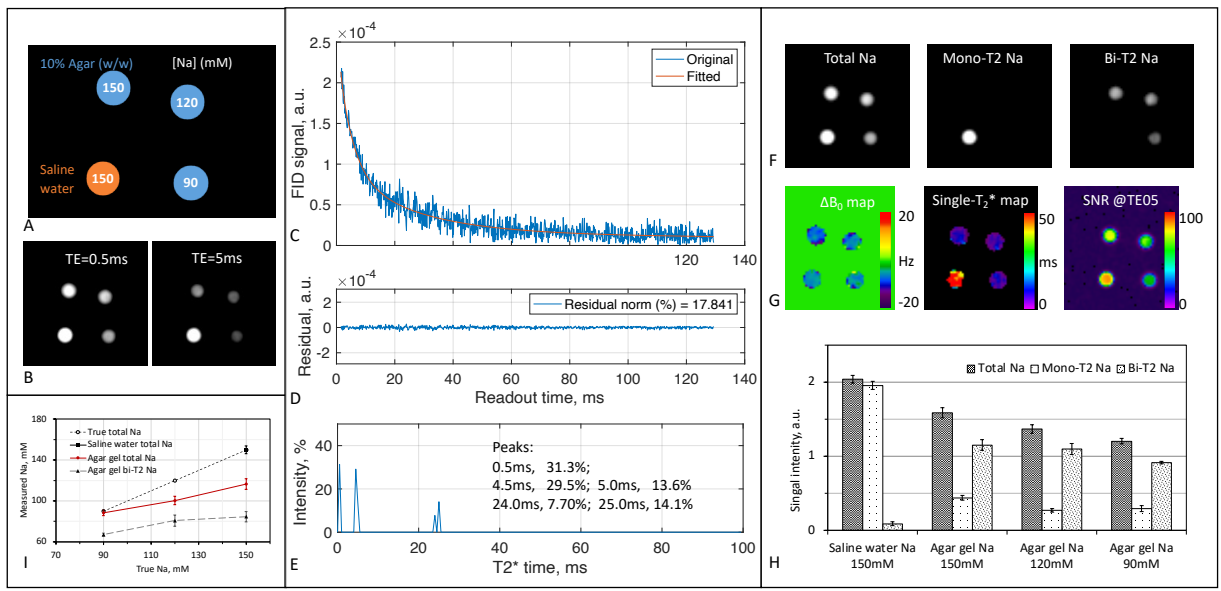

**Fig. E5. Phantom study. (A)** Phantoms of four tubes with sodium concentration: 150mM for the saline water (simulating mono-$T_2$ sodium ions) and 90, 120, 150 mM for the agar gels (simulating bi-$T_2$ sodium ions). **(B)** Sodium images of the phantoms at $TE_1/TE_2$=0.5/5ms, shown in the same window/level. **(C)** FID signals (original and fitted) from the four tubes (no averaging), with the correction for distortion at the first five data points. **(D)** Residual error of the fitting in C. **(E)** $T_2$* spectrum calculated from FID in C and used to produce the fitted FID. **(F)** Sodium images (total, mono-$T_2$, and bi-$T_2$) separated from the two images in B at ($T_2^*{}_{mo}$, $T_2^*{}_{bs}$, $T_2^*{}_{bl}$) = (50, 5, 25) ms according to E and G. **(G)** Maps of $\Delta B_0$ and single-$T_2$* calculated from the images in B, and a map of SNR at TE=0.5ms. **(H)** Separated sodium signals in the regions of tubes in F. **(I)** Quantified sodium concentration from F.

Fig. E5 summarizes outcomes of the phantom studies (*30*). Fig. E5A shows phantom arrangement of the four tubes and Fig. E5B are sodium images at $TE_1/TE_2$=0.5/5ms. Figs. E5C-E demonstrates FID signal from all the four tubes at averages=1, residual fitting error, and $T_2^*$ spectrum. Fig. E5F includes the sodium images separated at $(T_{2,mo}^*, T_{2,bs}^*, T_{2,bl}^*) = (50, 5, 25)$ ms according to $T_2^*$ spectrum in Fig. E5E. The maps of $\Delta B_0$, single-$T_2^*$, and SNR are in Fig. E5G. Signal intensity of the separated sodium signals (mean±SD) in the tubes is in Fig. E5H and the quantification of sodium concentration is in Fig. E5I. The separation in Fig. E5H recovered 95.8% of mono-$T_2$ sodium signal in the saline water tube, while leaving 4.2% to bi-$T_2$ sodium signal, much better than 20% left by the subtraction approach (*25*). The separation recovered 72.5, 80.4, and 75.9 % of bi-$T_2$ sodium signal in the agar tubes at sodium concentrations of 150, 120, and 90mM, respectively. The quantification of sodium concentration in Fig. E5I, when calibrated at the saline water, showed a systematic bias in total and bi-$T_2$ sodium concentrations, leading to an underestimate of sodium concentrations.

*Validation via the estimates of upper-limit for extra- and intracellular volume fractions*

Different from the phantom studies above, where ground truth of sodium concentrations is known, the human studies lack the ground truth for validation. Although each step in the MSQ separation is scientifically reasonable and its outcome should be believed correct, we alternatively and indirectly validate it by applying the separated sodium images to estimation of the upper-limit of extra- and intracellular volume fractions for which accurate measurements are available in literatures (*8, 26, 35, 36*). We assigned all mono-$T_2$ sodium to extracellular space and all bi-$T_2$ sodium to intracellular space, although they may co-exist in both extra- and intracellular spaces. The estimates (mean and SD) were made in ROIs of gray matter (GM) and white matter (WM) across three continuous slices via Eqs. (E4–E6) using sodium concentrations $C_{ex}$=145mM for extracellular space and $C_{in}$=15mM for intracellular space.

$$v_{ex} = 1/(1 + a) \tag{E4}$$
$$v_{in} = a/(1 + a) \tag{E5}$$
$$a = m_{bi}C_{ex}/m_{mo}C_{in} \tag{E6}$$

Fig. E6 presents these upper-limit estimates in the healthy group and shows significant (*P*=0.023) difference in volume fraction between the gray and white matters: 89.6±4.5 vs. 94.0±2.6 % for intracellular space, in line with the literature values of 75 vs. 92 % (*8, 26*); and 10.4±4.5 vs. 6.0±2.6 % for extracellular space, comparable to literature values of 14.1±1.8 (*35*) vs. 18±5 % (*36*).

No significant difference (*P*=0.953) was found between the healthy and patient groups (*N*=9 and 6, respectively).

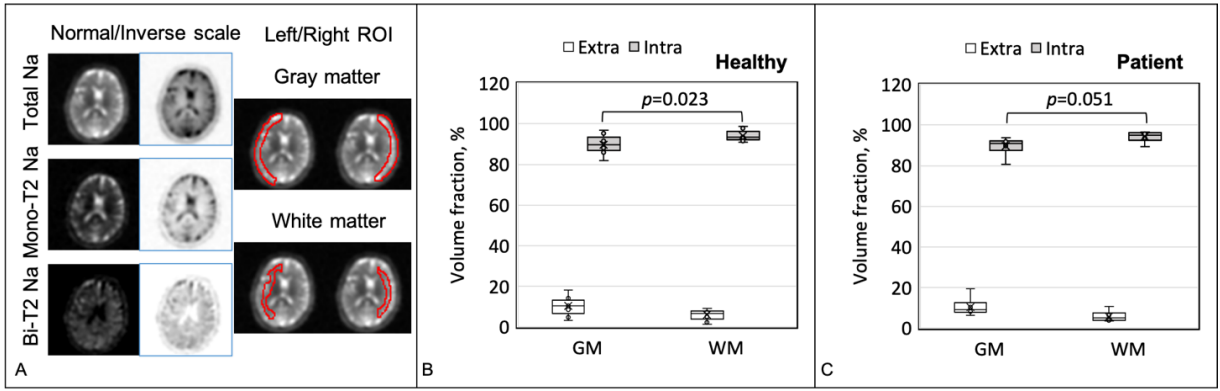

**Fig. E6. Estimates of upper-limit of the volume fraction for extra- and intracellular spaces. (A)** Typical sodium images (total, mono-T$_2$, and bi-T$_2$) and regions of interest (ROIs) for the gray matter (GM) and white matter (WM) in an axial slice from a healthy subject. **(B)** Volume fractions for the healthy group (*N*=9). **(C)** Volume fractions for the patient group (*N*=6). **Note:** difference in volume fraction is statistically significant between the gray and white matters (*P*=0.023) for the healthy group, but not for the patient group (*P*=0.051). No significant difference was observed between the healthy and patient groups for gray or white matter (*P*=0.953).

*Calculation of the separation sensitivity to T$_2$* values*

We used the most common set of T$_2$* values in our human studies, $\{T_{2,mo}^*, T_{2,bs}^*, T_{2,bl}^*\}$ = {50.0, 3.5, 15.0}ms, as true values, and calculate via Eqs. (1–3) the separation error $\{\delta m_{mo}, \delta m_{bi}\}$ relative to the ground truth of mono- and bi-T$_2$ sodium signals $\{m_{mo}, m_{bi}\}$ when adding errors $\delta T_{2,q}^*$, *q*=*mo, bs, bl*, up to ±20% to the true T$_2$* values. The relationship between $\{\delta T_{2,q}^*; q=mo, bs, bl\}$ and $\{\delta m_{mo}, \delta m_{bi}\}$ was plotted out in Extended Data Fig. E2. To focus on the relation of "$\delta T_2^* - \delta m$", TEs were sampled in an ideal case with TE$_0$= 0, ΔTE=1.0ms, and 80 TEs covering entire T$_2$* decays.

*Calculation for optimal number of TEs*

The calculations were implemented via Eq. (6) for three cases: an ideal case serving as reference, practical case #1 having a large number of TEs, and practical case #2 having a small number of TEs. The ideal case has 80 TEs, i.e., TE = 0, 1, 2, …,79ms, to cover entire range of T$_2$* decays. The two practical cases, suggested by existing human studies (*25, 31, 45, 46*) and being the most sensitive to T$_2$* decays (*45*), were investigated under the constraint of total scan time (TA) = 22min.

The practical case #1 has 8 TEs, i.e., TE = 0.5, 1, 2, 3, 4, 5, 7, and 10ms; and the practical case #2 only has two TEs, i.e., TE = 0.5 and 5.0ms with 4 averages for each TE.

*Numerical simulation*

These simulations were performed in MATLAB on the MacBook Pro laptop or Windows desktop, otherwise specified. Random noise of Gaussian distribution was generated using MATLAB function *randn(n)*, while the NNLS algorithm was implemented using [*x*, *resnorm*, *residual*] = *lsqnonneg(C, d)*.

*Phantom experiments*

We custom-built four phantoms, which were detailed in our previous work (*25*). The phantoms are 50-mL centrifuge tubes filled with a mixture of distilled water, 10% w/w agar powder, and sodium chloride (NaCl) at three concentrations (90, 120, and 150 mM) and at 150mM without agar, mimicking bi- and mono-$T_2$ sodium signals in brain tissues. MRI data acquisition was implemented using an SNR-efficient, three-dimensional (3D) pulse sequence called twisted projection imaging (TPI) (*31*), with parameters: rectangular RF pulse duration = 0.8ms, flip angle=80º (limited by SAR and TR), field of view (FOV)=220mm, matrix size=64, nominal resolution=3.44mm (3D isotropic), TPI readout time=36.32ms, total TPI projections=1596, TPI *p*-factor=0.4, TR=100ms, $TE_1/TE_2$= 0.5/5ms, averages=4, and TA=10.64min per TE-image. The sodium images were offline reconstructed on a desktop computer (OptiPlex 7050, 8GB memory, Windows 10, DELL, Round Rock, TX) using the gridding algorithm (*47, 48*) and a custom-developed programs in C++ (MS Visual Studio 2012, Microsoft, Redmond, WA). The mono- and bi-$T_2$ sodium signals was separated using a custom-developed program as described above.

# Supplementary Information for

**Multi-TE Single-Quantum Sodium ($^{23}$Na) MRI: A Clinically Translatable Technique for Separation of Mono- and Bi-T$_2$ Sodium Signals**

Yongxian Qian* *et al.*
.

*Corresponding author.  Email: Yongxian.Qian@nyulangone.org

**This PDF file includes:**

    Supplementary Text
    Figs. S1 to S5
    References (no)

*Extrapolation of N-term exponential decay*

We once accidently read a reference in literature about this topic, but could not find the citation at hand, thus summarize here the algorithm in our own language, specifically for the recovery of FID signals. If a signal $f(t)$ is an *N*-term exponential decay as defined in Eq. (S1) with parameters $\{A_i, b_i ; i = 1, 2, ..., N\}$, and is sampled at a uniform interval $\Delta t$, then a sample $f(t_0)$ at time $t_0$ can be represented by a linear combination of its late-time neighboring samples $\{f(t_0 + j\Delta t), j = 1, 2, ..., M\}$, as shown in Eq. (S2), with coefficients $\{a_j, j = 1, 2, ..., M \geq N\}$ to be determined.

$$f(t) = \sum_{i=1}^{N} A_i e^{-t \cdot b_i} \tag{S1}$$

$$f(t_0) = \sum_{j=1}^{M} a_j f(t_0 + j\Delta t) \tag{S2}$$

*Proof.* Extending $f(t_0 + j\Delta t)$ in Eq. (S2) according Eq. (S1) gives

$$f(t_0) = \sum_{j=1}^{M} a_j \left[ \sum_{i=1}^{N} (A_i e^{-t_0 \cdot b_i})(e^{-j\Delta t \cdot b_i}) \right]$$

$$= \sum_{i=1}^{N} A_i e^{-t_0 \cdot b_i} \left( \sum_{j=1}^{M} a_j e^{-j\Delta t \cdot b_i} \right). \tag{S3}$$

Select time-invariant coefficients $\{a_j, j = 1, 2, ..., M > N\}$ to satisfy Eq. (S4), thus Eq. (S2) holds.

$$\sum_{j=1}^{M} a_j e^{-j\Delta t \cdot b_i} = 1, \text{ for } i = 1, 2, ..., N. \tag{S4}$$

*Note*. The descriptions above are for backward extrapolation in time and used in the recovery of FID signal. The forward extrapolation also holds if $\Delta t$ is replaced with $-\Delta t$ in Eqs. (S2–S4). To find the unknown coefficients $\{a_j, j = 1, 2, ..., M\}$, Eq. (S2), instead of Eq. (S4), is usually used on such a segment of $f(t)$ that it is not distorted and involves all the *N* exponential decays. The number of data samples on the segment should be larger than *M* to form an over-determined problem in case of random noise existing in the signal $f(t)$.

*Correction for hardware-related distortion of FID signal*

Fig. S1 demonstrates a FID signal from a healthy subject (52 years old, male), with and without correction for the distortion at the first five ADC samples using Eq. (S2) with (*M*, *N*) = (5, 3). The correction removed distortion and reduced overall residual fitting error from 2.33% to 1.49%. The correction also improved resolution of short-$T_2^*$ components: from singlet at 2.5ms to doublet at 0.5ms and 2.5ms (Figs. S1A3, B3).

*Calculation stability of $T_2^*$ spectrum at a high resolution of $\Delta T_2^*$=0.5ms*

$T_2^*$ spectrum was calculated via Eq. (4) on an FID signal using an established algorithm – non-negative least squares (NNLS) – at a high spectral resolution of $\Delta T_2^*$=0.5ms in a range of 0.5–100ms. Such a high resolution raises a concern on stability of the calculation as the base functions at these spectral locations, $exp(-t/T_2^*)$, are not independent from each other. To address this concern, we employed singular value decomposition (SVD) to analyze the transfer matrix $\mathbf{E}$, and used numerical simulation to detail the impact of random noise on the T$_2$* spectrum.

The SVD analysis on the transfer matrix E are as the follows.

$$E_{i,j} \equiv exp(-t_i/T_{2,j}^*), \quad i = 1,2,\ldots,N, \quad j = 1,2,\ldots,M, \quad N \gg M \tag{S5A}$$

$$\mathbf{E^T E} = \mathbf{U \Sigma V^T} \tag{S5B}$$

$$\mathbf{\Sigma} = \text{diag}(\sigma_1, \sigma_2, \ldots, \sigma_M) \tag{S5C}$$

with sampling time $t_i = TE + (i-1) * \Delta t$ and spectral location $T_{2,j}^* = j * \Delta T_2^*$. Singular values $\{\sigma_j, j = 1,2,\ldots,M\}$ determine stability of the calculation for T$_2$* spectrum in terms of random noise interference in Eq. (4). Correlation coefficients between the base functions are also calculated.

$$R_{j1,j2} = (E^T E)_{j1,j2}/\sqrt{(E^T E)_{j1,j1}(E^T E)_{j2,j2}}, \quad j1, j2 = 1,2,\ldots,M \tag{S6}$$

At $\Delta t = 0.05$ms, $TE$=0.2ms and $N$=2048, the singular values and correlation coefficients were calculated and shown in Fig. S2. The singular value $\sigma$ quickly decreases to zero (<10$^{-10}$) at index (15, 13, 10, 9) when $\Delta T_2$* increases from 0.5ms to 1.0, 3.0 and 5.0ms, respectively. This indicates the existence of null subspace or multiple solutions for T$_2$* spectrum (Fig. S2, top). The normalized correlation coefficients $R$ between any two T$_2$* base functions is spreading out from diagonal line, confirming non-orthogonal between the base functions (Fig.S2, bottom). However, the extent of spreading is narrower for short T$_2$* values at high resolution $\Delta T_2$* = 0.5ms than at low resolution $\Delta T_2$* = 5ms.

Numerical simulation for the impact of random noise on the T$_2$* spectrum was performed at three popular components, T$_2$*=3, 15, and 50ms with relative amplitudes A=30, 20, and 50, respectively, plus an additive normal random noise generated by function *randn(n,1)*, at SNR $\equiv f(t$=0)/SD = 100, 50, and 25. Outcomes of the simulations were summarized in Fig. S3, where peak parameters at doublets (Fig. S3, bottom) were linearly combined with amplitude-weighting by left- and right-peaklets in Eq. (S7). The best spectrum was achieved at SNR=100 among the three noisy cases, relative to no noise.

$$T_2^* = (A_L * T_{2,L}^* + A_R * T_{2,R}^*)/A \tag{S7A}$$

$$A = A_L + A_R \tag{S7B}$$

### *Measurement stability of FID signals on whole brain: B$_0$ shimming*

The $B_0$ shimming may change from subject to subject in routine practice, leading to a concern on measurement stability of the FID signals, thus the $T_2^*$ spectra, from whole brain across subjects. This concern is addressable because sodium ($^{23}$Na) MRI has about 4-fold lower resonance frequency than proton ($^1$H) MRI (e.g., 33.8 vs. 127.7 MHz at 3T), and the manual shimming (three iterations) is better than auto shimming. Fig. S4 shows the results of all 15 subjects studied, with a small standard deviation (SD) in whole-brain histograms. There was no significant difference between the healthy and patient groups (*P*=0.908). Thus, the manual shimming, or $\Delta B_0$, is stable.

*Invisibility of CSF $T_2^*$ peaks in the spectrum: single $T_2^*$ mapping*

Cerebrospinal fluid (CSF) in the brain is known to have a $T_2^*$ value of ~50ms as seen in single-$T_2^*$ maps (Figs. 7, 8). But this sodium population was not observed in the $T_2^*$ spectra. This phenomenon might be caused by small volume of CSF relative to whole brain. To confirm this cause, Fig. S5 presents two representative whole-brain histograms of single-$T_2^*$ mapping, with very small numbers of voxels (invisible bins) for CSF at $T_2^*$ ~50ms.

**Fig. S1. FID signal and T$_2$* spectrum with and without correction.**

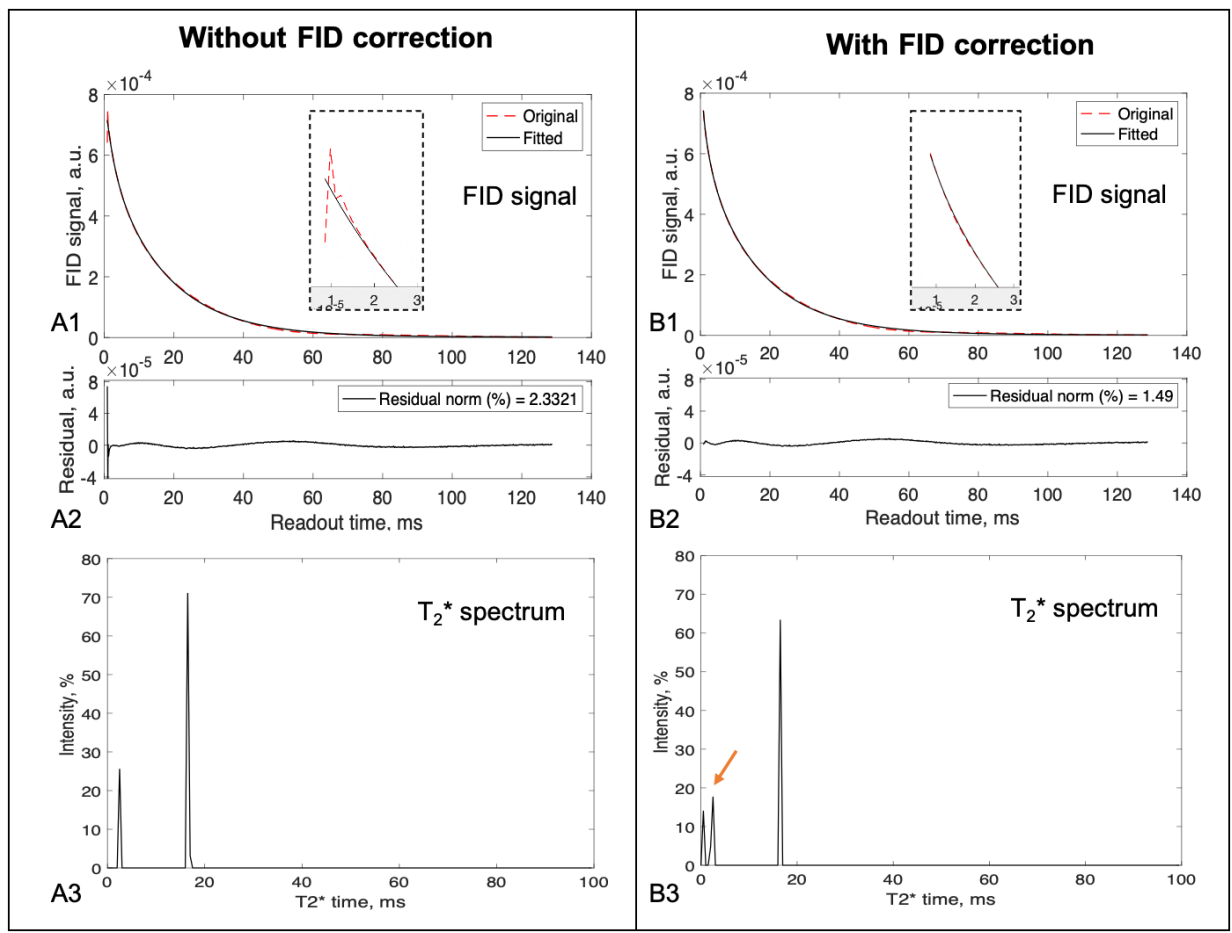

**Fig. S1.** FID signals (top) and T$_2$* spectra (bottom) from whole brain of a healthy subject (52 years old, male), with (Fig. S1B) and without (Fig. S1A) correction for FID distortion at the first five samples shown in the insets. In the middle are residual errors from the fitting using the T$_2$* spectra in the bottom. The FID correction removed the distortion, significantly reduced residual error, and clearly improved resolution of short-T$_2$* components from singlet at 2.5ms to doublet at 0.5ms and 2.5ms as well as peaks' intensity (Fig. S1B3). Data acquisition: 3T scanner (Prisma, Siemens) with a custom-built dual-tuned ($^1$H-$^{23}$Na) 8-channel head array coil (*32*), *fid* sequence, rectangular RF duration=0.5ms, TE/TR=0.35/300ms, averages=128, ADC samples=1024 at an interval of 0.125ms.

**Fig. S2. Calculation stability of T₂* spectrum.**

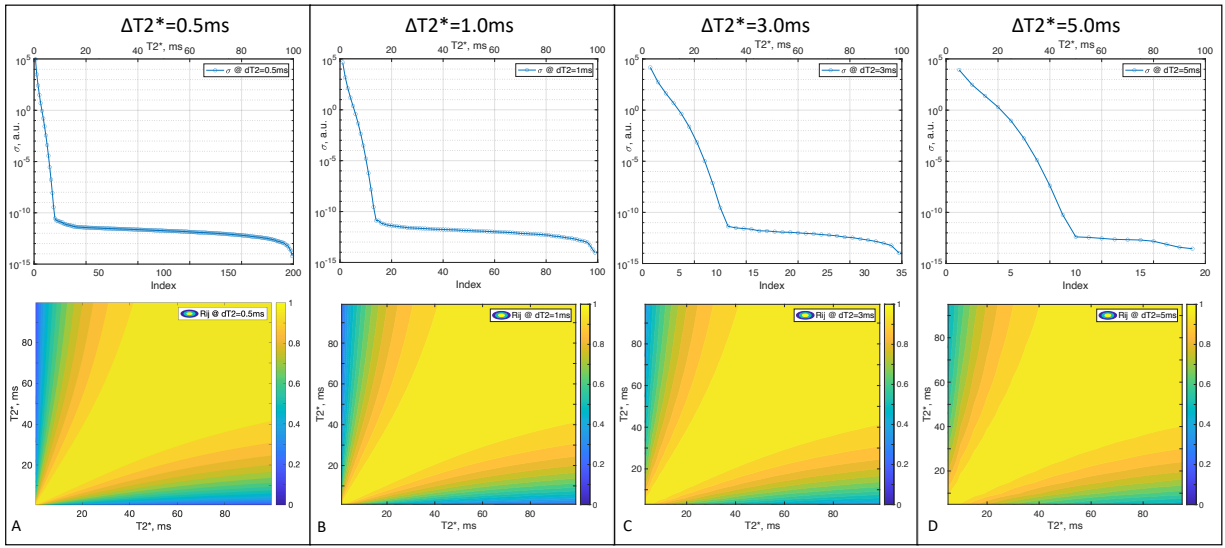

**Fig. S2.** The SVD singular values of matrix $E^T E$ (top) and correlation coefficients of the base-in matrix $E$ (bottom). **(A) – (D)** T$_2$* spectral resolution at ΔT$_2$* = 0.5, 1.0, 3.0, and 5.0ms. In the top, singular value $\sigma$ quickly decreases to zero (<10$^{-10}$) at index (15, 13, 10, 9) respectively, indicating the existence of null subspace or multiple solutions for the T$_2$* spectrum. In the bottom, the normalized correlation coefficient $R_{j1,j2}$ between any two T$_2$* base functions exp(-$t/T_{2,j}$*) is spreading out from diagonal line, showing non-orthogonal between the base functions.

**Fig. S3. Numerical simulation for the impact of random noise on $T_2^*$ spectrum.**

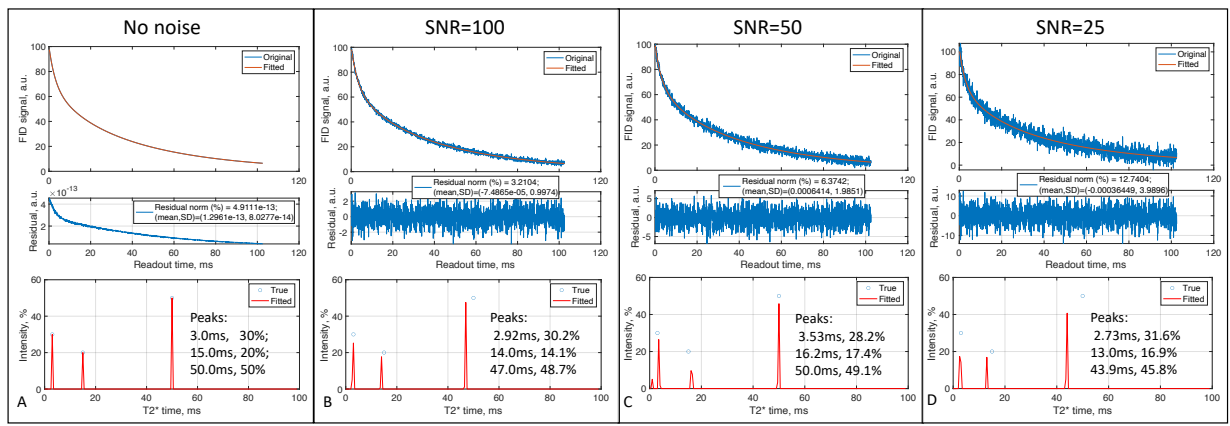

**Fig. S3.** Calculation stability of $T_2^*$ spectrum using the algorithm NNLS via MATLAB function *lsqnonneg(C,d)* and the numerical simulations at three popular components: $T_2^*$=(3, 15, 50) ms with relative amplitudes A=(30, 20, 50) plus an additive random noise generated by function *randn(n,1)*. **(A) – (D)** are the simulations at $\Delta T_2^* = 0.5$ms with noise at three typical values SNR = 100, 50, and 25. The peak parameters at doublets (bottom) were linearly combined with amplitude-weighting (Eq. S7). The best spectrum was achieved at SNR=100 among the three noisy cases.

# Fig. S4. Measurement stability of FID signals on whole brain: B$_0$ shimming

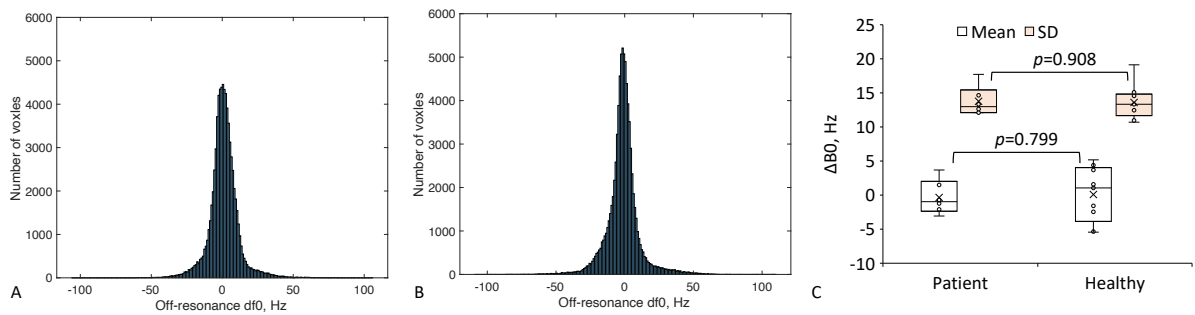

**Fig. S4.** Whole brain histograms of $\Delta B_0$ mapping at TE$_1$/TE$_2$= 0.5/5ms under a manual shimming procedure (3 iterations). **(A)** A representative histogram from a healthy subject (52 years old, male), with mean±SD = 1.0±10.7 Hz. **(B)** Another example from a patient with epilepsy (31 years old, male), with mean±SD = -1.2±12.1 Hz. **(C)** Mean and SD distribution of whole-brain $\Delta B_0$ histograms from all 15 study subjects including 9 healthy and 6 patients, showing no significant difference between the two groups (healthy vs. patient), $P = 0.799$ for the mean and $P = 0.908$ for the SD.

**Fig. S5. Invisibility of CSF T$_2$* peak in the spectrum: single T$_2$* mapping.**

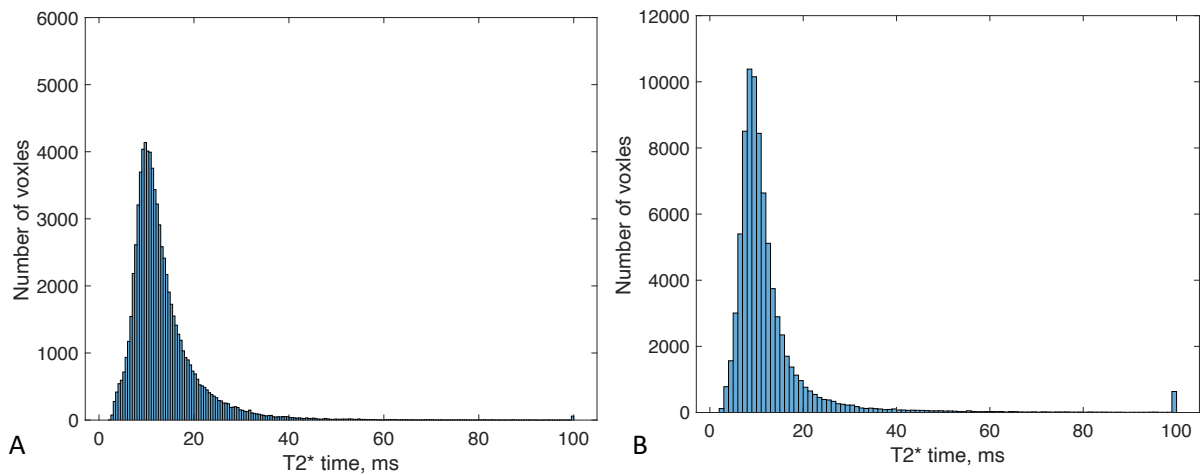

**Fig. S5.** Representative whole-brain histograms of single-T$_2$* mapping at TE$_1$/TE$_2$= 0.5/5ms. **(A)** A healthy subject (52 years old, male). **(B)** An epilepsy patient (31 years old, male). These, as well as the other healthy subjects and patients we studied, showed very small numbers (invisible bins) of voxels for CSF at T$_2$* ~ 50ms. Note: a visible bin at T$_2$*=100ms counts voxels of T$_2$* values >= 100ms.